\documentclass[copyright,creativecommons]{eptcs}
\providecommand{\event}{AUTOMATA \& JAC 2012} 

\usepackage{stmaryrd,amssymb,graphicx,url}
\urldef{\mailsc}\path|imai@iec.hiroshima-u.ac.jp|

\newcommand{\rmk}{\textbf{Remark: }}
\newcommand{\NN}{\mathbb{N}}
\newcommand{\ACA}{\mathcal{A}}
\newcommand{\QQ}{\mathcal{Q}}
\newcommand{\CC}{\mathfrak{C}}
\newcommand{\RR}{\mathbb{R}}

\newtheorem{definition}{Definition}

\title{A Universal Semi-totalistic Cellular Automaton on Kite and Dart Penrose Tilings}
\def\titlerunning{A Universal CA on Penrose Tilings}
\author{Katsunobu Imai
\institute{Graduate School of Engineering, Hiroshima University}
\email{imai@iec.hiroshima-u.ac.jp}
\and
Takahiro Hatsuda
\institute{Graduate School of Engineering, Hiroshima University}
\email{thatsuda@iec.hiroshima-u.ac.jp}
\and
Victor Poupet
\institute{Laboratoire d'Informatique Fondamentale de Marseille, Aix-Marseille University}
\email{victor.poupet@lif.univ-mrs.fr}
\and
Kota Sato
\institute{Graduate School of Engineering, Hiroshima University}
\email{ksatou@iec.hiroshima-u.ac.jp}
}
\def\authorrunning{K. Imai, T. Hatsuda, V. Poupet\& K. Sato}

\begin{document} 
\maketitle

\begin{abstract}
	In this paper we investigate certain properties of semi-totalistic cellular automata (CA) on the well known quasi-periodic \emph{kite and dart} two dimensional tiling of the plane presented by Roger Penrose.
	
	We show that, despite the irregularity of the underlying grid, it is possible to devise a semi-totalistic CA capable of simulating any boolean circuit and any Turing machine on this aperiodic tiling.

\end{abstract}

\section{Introduction}

Since the discovery of quasi-periodic tilings and the related quasi-cristals, several researchers have studied the behaviors of cellular automata (CA) on quasi-periodic tilings. 

In 2005, Chidyagwai and Reiter showed that the broken symmetry of quasi-periodic tilings, while still retaining a highly organized structure, could be used to simulate complex growth of snow crystals~\cite{CR05} by studying real-valued CA on such lattices. They could produce global $n$-fold symmetry models where regular hexagonal grids could only produce $6$-fold symmetry models.


Reiter also investigated the behaviors of cyclic CA rules on several quasi-periodic tilings. Cyclic CA are special CA that can be regarded as a model of excitable media~\cite{Dewdney89,Fisch90}. Once a cell enters a specific state, it must go through a fixed cycle of other states before being active again, thus implementing a \emph{refractory period} mechanism. Reiter found that cyclic CA on quasi-periodic tilings, despite having a very simple local behavior, can take advantage of the lattice geometry to exhibit complex and smooth spiral patterns~\cite{Reiter10}. A variety of Cyclic CA rules are illustrated and commented in \cite{Wojtowicz}. 

Variants of John H.~Conway's \emph{Game of Life}~\cite{Gardner70} have also been studied on Penrose tilings. In particular, Owens and Stepney found many stable and periodic patterns, similar to the ones known in the usual euclidian \emph{Game of Life}~\cite{OS08,OS10}. Despite this, no equivalent to the \emph{glider} structures of the euclidian case are known on quasi-periodic variants of the \emph{Game of Life}. Indeed, because of the non-periodic nature of the underlying grid, the very definition of gliders is unclear.

In this paper, we investigate some semi-totalistic CA on Penrose's \emph{kite and dart} aperiodic tiling of the plane. We show in particular that there exists an $8$-states semi-totalistic CA capable of simulating any boolean circuit. The usual difficulty of synchronizing signals, made even more problematic in an aperiodic setting, is avoided by embedding universal asynchronous logical elements. This construction is similar to that employed to construct a universal CA on an hyperbolic space~\cite{IIM06}. Once we are able to simulate logical circuits, we use Morita's construction~\cite{Morita10} to simulate a Turing machine with simple finite memory logical elements.

\section{Semi-totalistic Cellular Automata on Penrose Tilings}

\subsection{Penrose's \emph{Kite and Dart} Tiling}

This famous tiling of the plane was discovered by Roger Penrose as a simplification of his previous 6-tiles aperiodic tilings. This tiling is generated by two quadrilateral proto-tiles~: a convex one called \emph{kite} and a non-convex one called \emph{dart}. To avoid periodic arrangements, tiling constraints are added on the tiles~\cite{Gardner77}.

This tiling has been widely studied, and its many properties will not be discussed in details in this article.

There exist infinitely many different possible tilings of the plane with the \emph{kite} and \emph{dart} tiles. They are all aperiodic but quasi-periodic, that is to say that there exists a constant $k\in\RR$ such that for any $r\in\RR$, any finite pattern of radius $r$ that appears on the tiling appears in every area of radius $k\times r$.

\subsection{The Moore Neighborhood on a Penrose Tiling}

In order to define a CA on a Penrose tiling, we need to define a neighboring relation between the different tiles. To do so, we extend the usual Moore neighborhood (8 closest neighbors on the regular square grid) by considering that two Penrose tiles are neighbors if they share at least one vertex.

\begin{figure}[ht]
  \begin{center}
    \includegraphics[width=10cm]{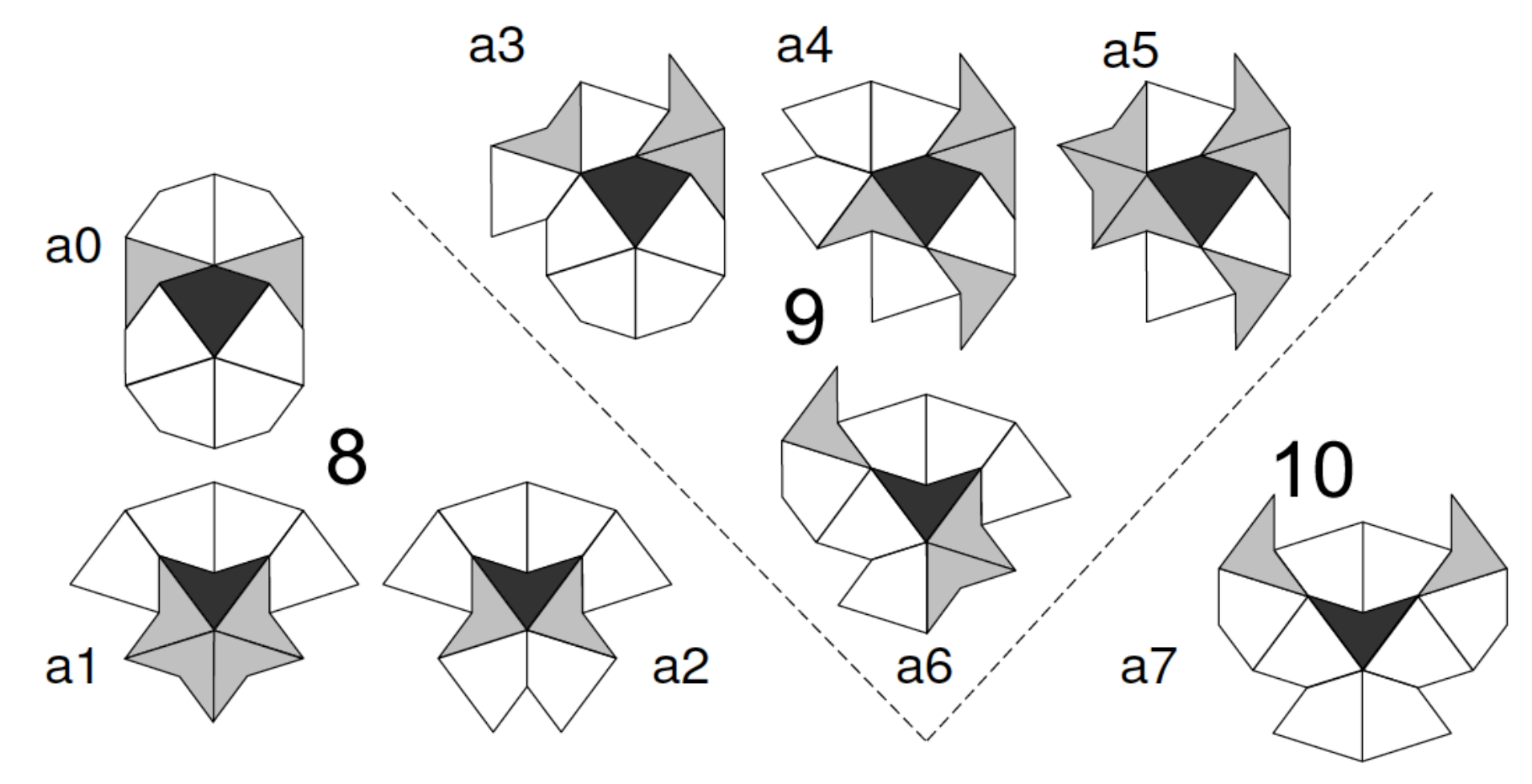}
    \caption{The generalized Moore neighborhoods on a \emph{kite} and \emph{dart} Penrose tiling~\cite{OS08}}
    \label{fig:kd-moore-neighborhood}
  \end{center}
\end{figure}

Because all tiles are not identical and not periodically arranged, different tiles might have different neighborhood. In the case of the \emph{kite and dart} tiling, there are 8 different possible neighborhoods~\cite{OS08}, illustrated by Figure~\ref{fig:kd-moore-neighborhood}.

It is important to notice that not only these neighborhoods have different shapes, they also have different sizes ranging from a minimum of 9 cells to a maximum of 11. Although this would normally be a problem when defining a cellular automaton (because it would not be possible to define a unified rule for all cells), we will see that the special class of semi-totalistic CA is especially well suited for non-homogeneous neighborhoods.

\subsection{Semi-totalistic Cellular Automata}

Because we want to work on aperiodic tilings of the plane, the usual definition of a cellular automaton must be changed in order to be compatible with the fact that different cells might have different neighborhoods. The subclass of semi-totalistic cellular automata is particularly well suited for this as the position of the different neighbors is irrelevant, only the number of neighbors in a given state has an effect on the evolution of a cell.

\begin{definition}[Semi-totalistic Cellular Automaton]
	A \emph{semi-totalistic cellular automaton} is a pair $\ACA = (N, \delta)$ where~:
	\begin{itemize}
		\item $N\in \NN$ is the number of states of the automaton. The set of states of the automaton is therefore $\QQ_\ACA = \{0, 1, \ldots, N-1\}$~;
		\item $\delta : \QQ_\ACA\times \NN^N\rightarrow \QQ_\ACA$ is the totalistic local transition function of the automaton.
	\end{itemize}
\end{definition}

\begin{definition}[Configuration]
	A \emph{configuration} of the automaton $\ACA = (N, \delta)$ over a tiling of the plane by Penrose's tiles is a mapping $\CC$ that associates a state of $\QQ_\ACA$ to each tile.
\end{definition}

In accordance with usual cellular automaton terminology, in this context, the tiles of the Penrose tiling will be denoted as \emph{cells}.

\begin{definition}[Global Evolution]
	Given a configuration $\CC$, the image $\CC'$ of $\CC$ by the automaton $\ACA=(N, \delta)$ is such that for each cell $c$, the state $\CC'(c)$ is defined by
	\[ \CC'(c) = \delta(q_c, n_0, n_1, \ldots, n_{N-1})\]
	where $q_c$ is the current state of $c$ and for every $i\in\llbracket 0, N-1\rrbracket$, $n_i$ is the number of neighbors of $c$ in state $i$.
	
	We will denote by $\ACA(\CC)$ the image of a configuration $\CC$ by such a process. The global action of the automaton can be iterated over each new configuration, defining in such a way the infinite evolution of the initial configuration.
\end{definition}

In other words, at each time every cell of the automaton updates its state according to the number of neighbors it has in each state.

\rmk In order to be as general as possible, and to be able to define a CA independently of the lattice on which we will consider it, we have defined the local transition function as a function from and infinite set $\QQ_\ACA\times \NN^N$. However, once the geometric structure of the cells has been decided (in our case the Penrose tiling), it is sufficient to define the local transition function over the tuples that can actually appear. In particular, since we know that in the \emph{kite and dart} tiling all cells have between 9 and 11 neighbors (themselves included), only the input tuples of the form $(q, n_0, n_1, \ldots, n_{N-1})$ where $9 \leq \sum_{i}n_i\leq 11$ should be considered.
\subsection{Previous Studies}

Owens and Stepney studied the \emph{Game of Life} on Penrose tilings~\cite{OS08,OS10}.

Even though the average size of the neighborhoods is larger than that of the 
standard Moore neighborhood in the regular case, the same parameters for the rule (a cell survives with 2 or 3 live neighbors and a new cell is born when surrounded by 3 live neighbors) exhibits behaviors very similar to the classical \emph{Game of Life} and many similar stable and periodic patterns were found.

However, no \emph{glider} or similarly propagating pattern has been found so far, possibly because of the strong irregularity of the encountered geometry as the signal moves. \emph{Gliders} being the heart of the computational and intrinsic universality proofs of the \emph{Game of Life} in the classical case, this approach cannot be used in the quasi-periodic case.

Another noteworthy difference between the Penrose case and the classical case is that long string-like (closed or not) connected alive cells can be stable on a Penrose tiling. Such patterns, called \emph{rings} and \emph{snakes} have been studied in~\cite{OS10}.

\section{A Universal CA on \emph{Kite and Dart} Penrose Tilings}

In this section we present the construction of a semi-totalistic CA working on Penrose's \emph{kite and dart} tilings that is logically universal in the sense that it can simulate the behavior of any boolean circuit.

\subsection{Universal Reversible Logical Elements}

In order to construct a universal CA, we will use very simple reversible logical elements with $1$-bit memory.

The usual difficulty when embedding circuits in a larger structure (such as a cellular automaton) is to synchronize the different signals. In the case of an aperiodic geometry, the problem is even harder as construction patterns cannot be precisely shifted in a given direction. To solve this problem we use asynchronous elements that can simulate any circuit while only having one signal at a time.

\begin{figure}[ht]
  \begin{center}
    \includegraphics[width=1.0\linewidth]{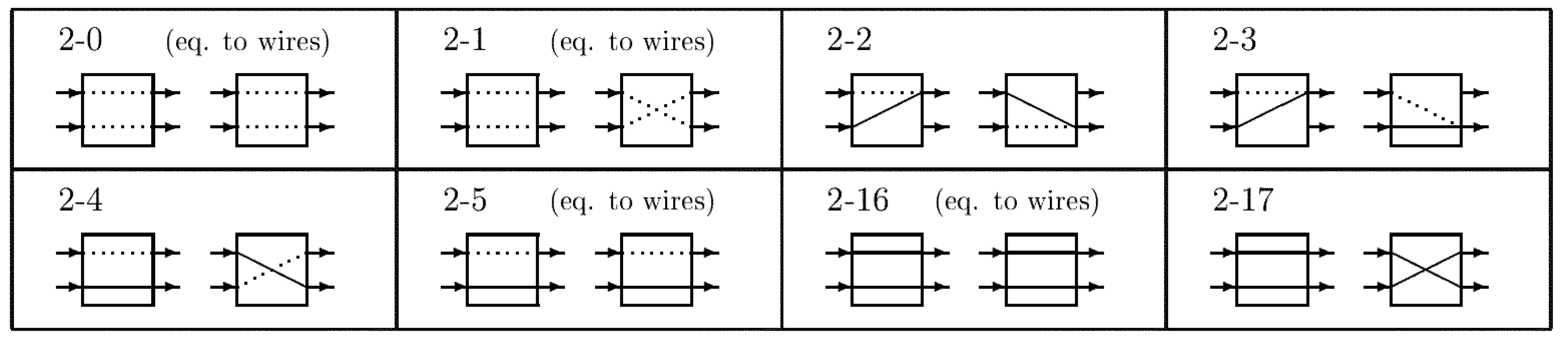}
    \caption{$2$-symbol reversible logical elements with $1$-bit memory~\cite{MOA11}.}
    \label{fig:reversible-elements}
  \end{center}
\end{figure}

Figure~\ref{fig:reversible-elements} shows every $2$-symbol reversible 
logical elements with $1$-bit memory. There are only eight types of such elements. The left part of each illustration represents the element in state 0, the right part being state 1. Each element only accepts one signal from one of two possible inputs (left side) at a time. Lines (full and dotted) represent the output through which the signal will exit the element. 

When a signal traverses the element through a full line, the state of the element is changed. If the signal traverses through a dotted line its state is unchanged.

It has been shown that any reversible Turing machine can be constructed by 4-symbol reversible logical elements ({\em rotary elements})~\cite{Morita10}. Its computing process is executed by a signal, thus there is no need to synchronize signals. The rotary element can be realized by any combination of two elements amongst 2-3, 2-4, and 2-17~\cite{LPAM08,MM11}. 
So the combination of two elements are universal. 

In our construction we will use elements 2-3 and 2-17.

\subsection{General Overview of the Automaton}

The automaton has 8 states~:
\begin{itemize}
	\item State 0 is a \emph{quiescent} state, meaning that a cell in state 0 surrounded by other cells in the same state will not change. This state is therefore used as a ``background'' for the whole construction and a ``filler'' in the local transition rule to account for the different sizes of neighborhoods (the number of other states in the neighborhood is significant but the exact number of 0 states is not).
	\item States 1, 2 and 3 are used to transmit the signal through wire-like structures. Connected cells in state 1 will act as a wire along which the signal moves. The signal itself is made of two states to orient it. State 3 is the head of the signal and state 2 is the refractory state that remains after a cell was in state 3 in order to avoid the signal going back through the wire in the opposite direction.
	\item Finally, states 4 to 7 are used internally in the logical elements to control the behavior and store the inner state of each element.
\end{itemize}

Any boolean circuit can be simulated by an asynchronous $1$-signal circuit containing only copies of the elements 2-3 and 2-17, connected by wires~\cite{LPAM08,MM11}. Because the circuit is two-dimensional, wires must also be able to cross.

It is therefore enough to show that our automaton can accurately simulate the behavior of the logical elements and that these elements can be connected by wires.

\subsection{Simple Wires}

Wires can be easily represented by a continuous line of neighboring cells in state $1$ surrounded by cells in state $0$. It is important that each cell of the wire is connected to exactly two other cells of the same wire. This constraint is not very restrictive though, and it is easy to draw a wire going from any point of the tiling to any other. Simple wires cannot cross, and we will need a specific construction to allow crossings.

A signal on the wire is represented by two consecutive cells in states 3 (front) and 2 (back).

The transition rules needed to describe the behaviors on the wires are quite simple~:
\begin{itemize}
	\item a cell in state 1 that has a neighbor in state 3 (head of the signal) changes to state 3~;
	\item a cell that is in state 3 changes to state 2 (refractory state)~;
	\item a cell in state 2 returns to the simple wire state 1~;
	\item finally, cells in state 1 remain in that state in the absence of a neighboring signal.
\end{itemize}

\begin{figure}[htbp]
  \begin{center}
    \includegraphics[width=0.8\linewidth]{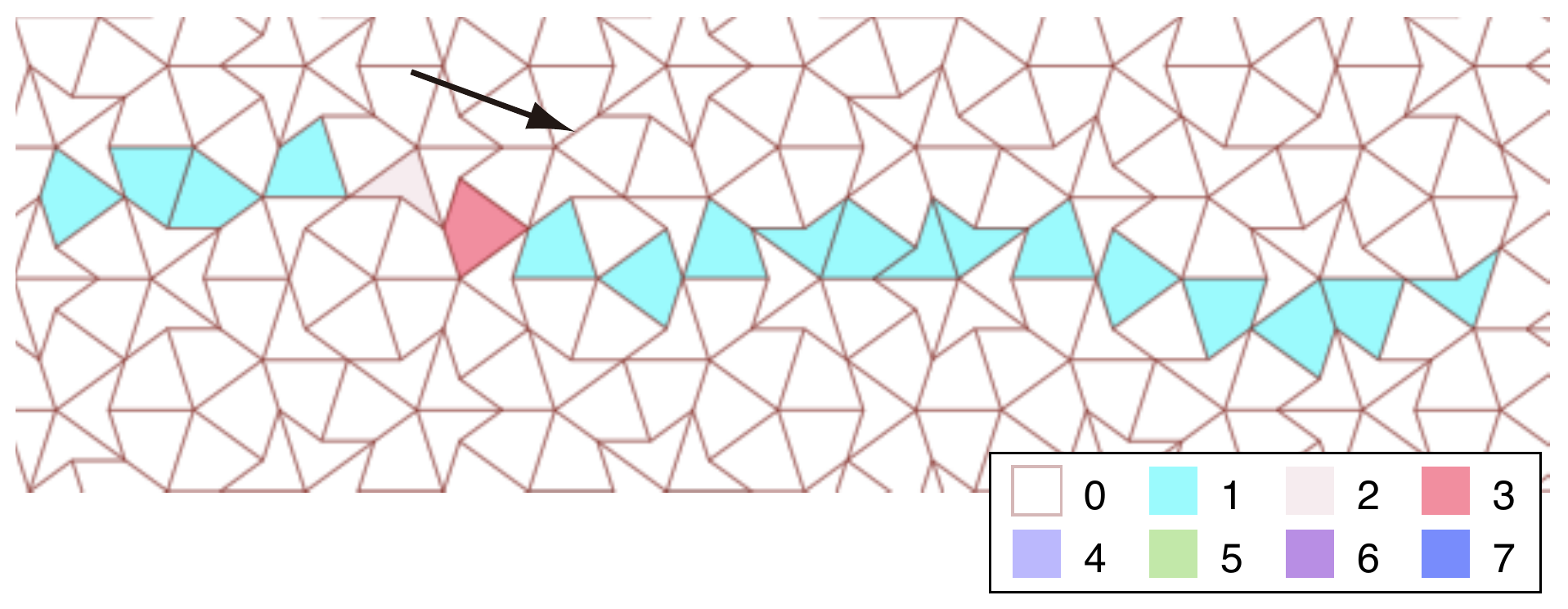}
    \caption{A signal on a wire and their state-color mapping table.}
    \label{fig:wire-and-signal}
  \end{center}
\end{figure}

Figure~\ref{fig:wire-and-signal} illustrates a wire on which there is a signal moving towards the right.

\subsection{Logical Elements}

The simulation of the elements 2-3 and 2-17 can be done by adding new states to the automaton. It would be very tedious to devise a construction that could behave as a specific logical gate in every possible neighborhood, but we only need to have a pattern that works in a specific geometric neighborhood because the tilings produced by the \emph{kite and dart} tiles are all quasi-periodic.

This means that if we can position the widget at a given position because the surrounding tiles correspond to what we need, there are such appropriate positions in every sufficiently large surface of the tiling.

\begin{figure}[htbp]
  \begin{center}
	\includegraphics[width=.8\linewidth]{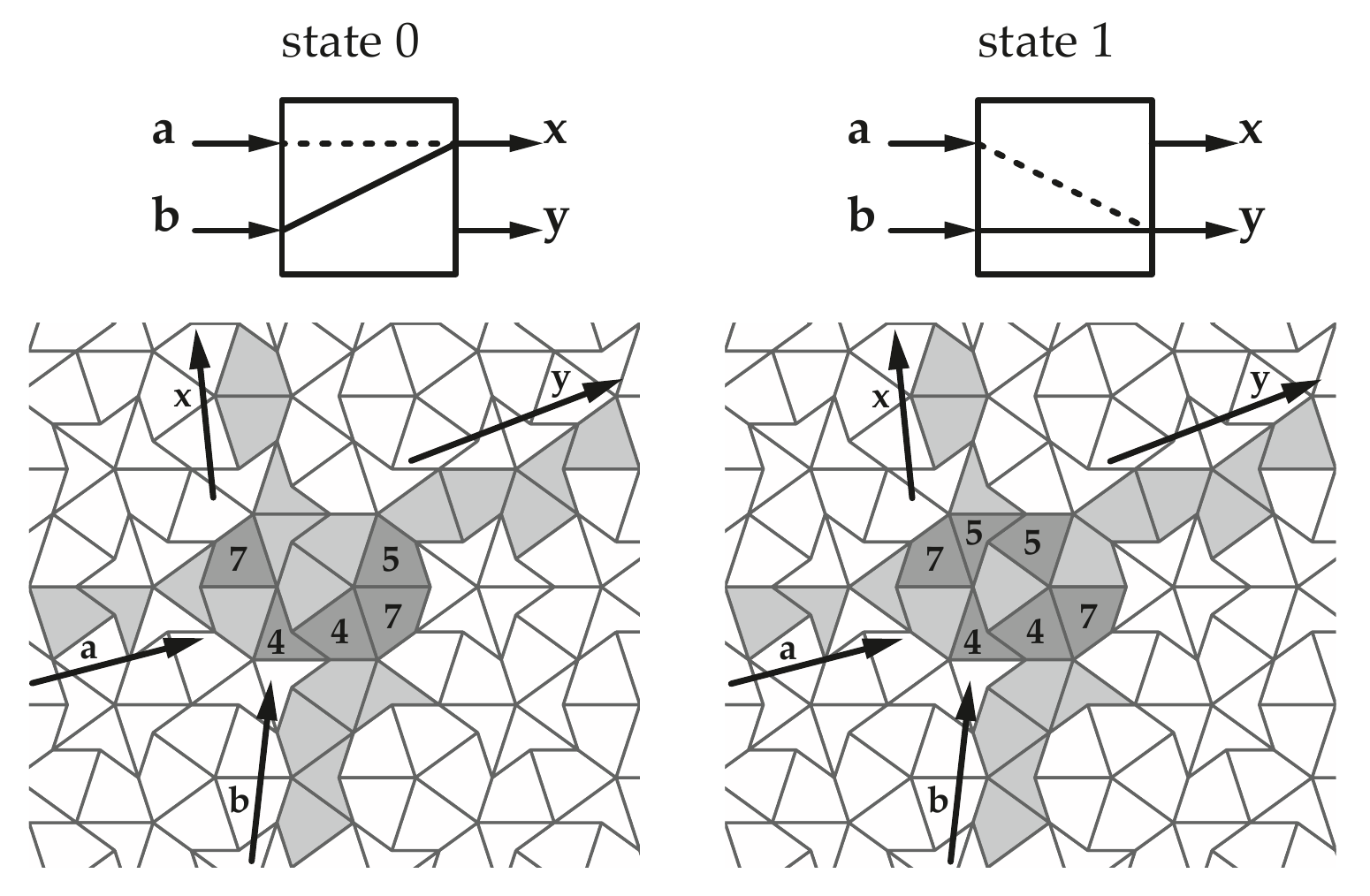}
    \caption{Logical element 2-3 in states 0 (left) and 1 (right). White cells in state 0, light gray cells in state 1.}
    \label{fig:2-2-3}
  \end{center}
\end{figure}

\begin{figure}[htbp]
  \begin{center}
    \includegraphics[width=0.8\linewidth]{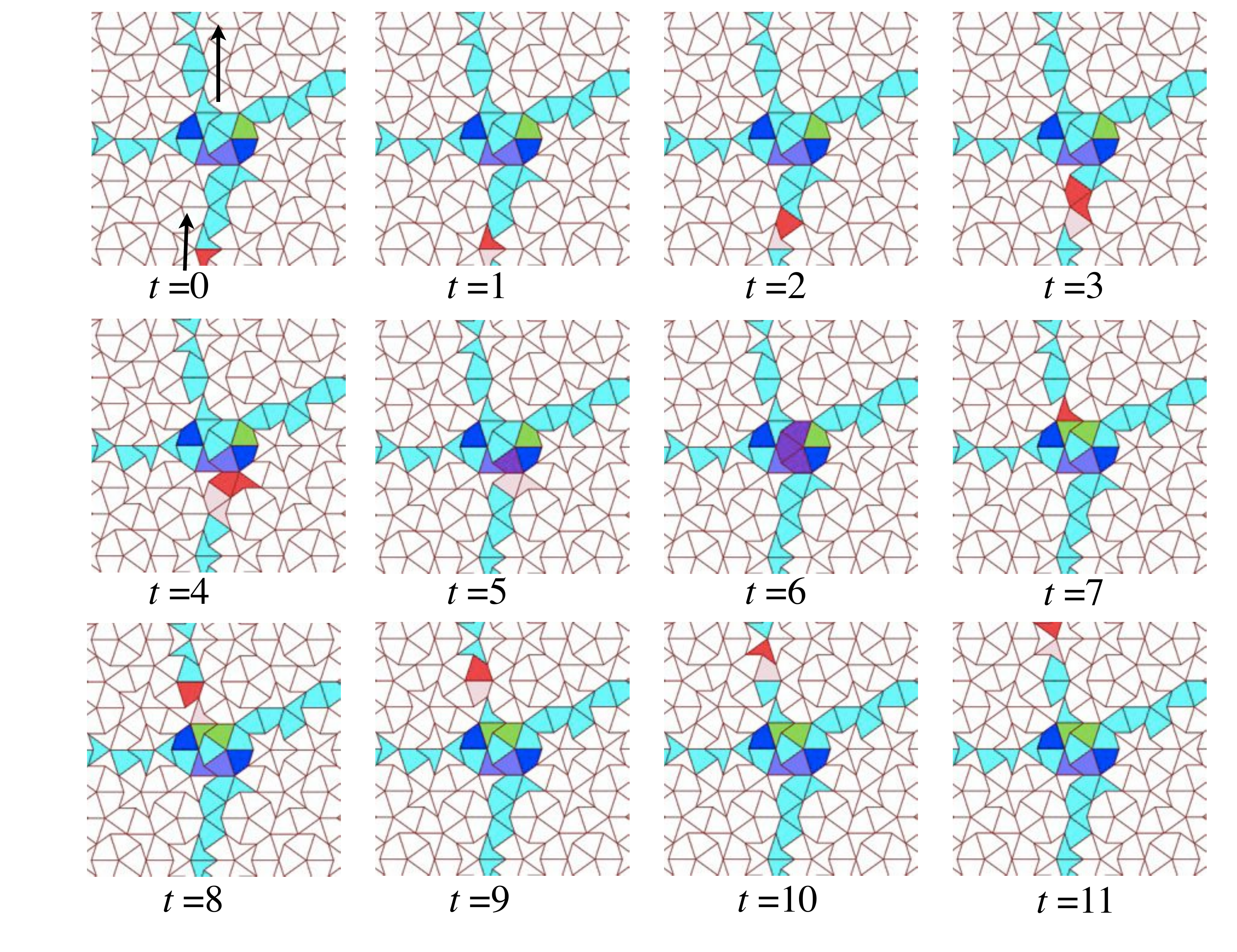}
    \caption{Element 2-3: a transition from state 0, input $b$ to state 1, output $x$.}
    \label{fig:2-3-0b-1x}
  \end{center}
\end{figure}

For instance, Figures~\ref{fig:2-2-3} and \ref{fig:2-3-0b-1x} illustrate a widget that behaves as element 2-3. This widget must be located on a neighborhood of type $a0$ (according to the classification of Figure~\ref{fig:kd-moore-neighborhood}) in order to work properly as any other positioning would change the number of neighbors of one of the cells and significantly alter its behavior.

But because of the quasi-periodicity of the tiling, there are such locations in every area of the tiling. We can therefore put different copies of the widget in different areas of the tiling and connect them with wires as needed (the wires do not require specific arrangements of the tiles).

The quasi-periodicity can also be used for larger constructions. If we create part of a circuit in some specific location and want to copy it somewhere else, we know that there exists another appropriate location not too far from the original (the maximal distance is proportional to the radius of the widget we need to copy). Because of this, it is enough to verify that the widget works as intended in its appropriate environment to ensure that we can use it to build arbitrarily large circuits.

Figure~\ref{fig:2-2-3} shows the two patterns that will simulate the element 2-3 with internal states 0 and 1 respectively. Figure~\ref{fig:2-3-0b-1x} illustrates the evolution of this pattern when a signal enters from its second input ($b$) and exits through its first output ($x$) while changing its internal state.

\begin{figure}[htbp]
  \begin{center}
	\includegraphics[width=.8\linewidth]{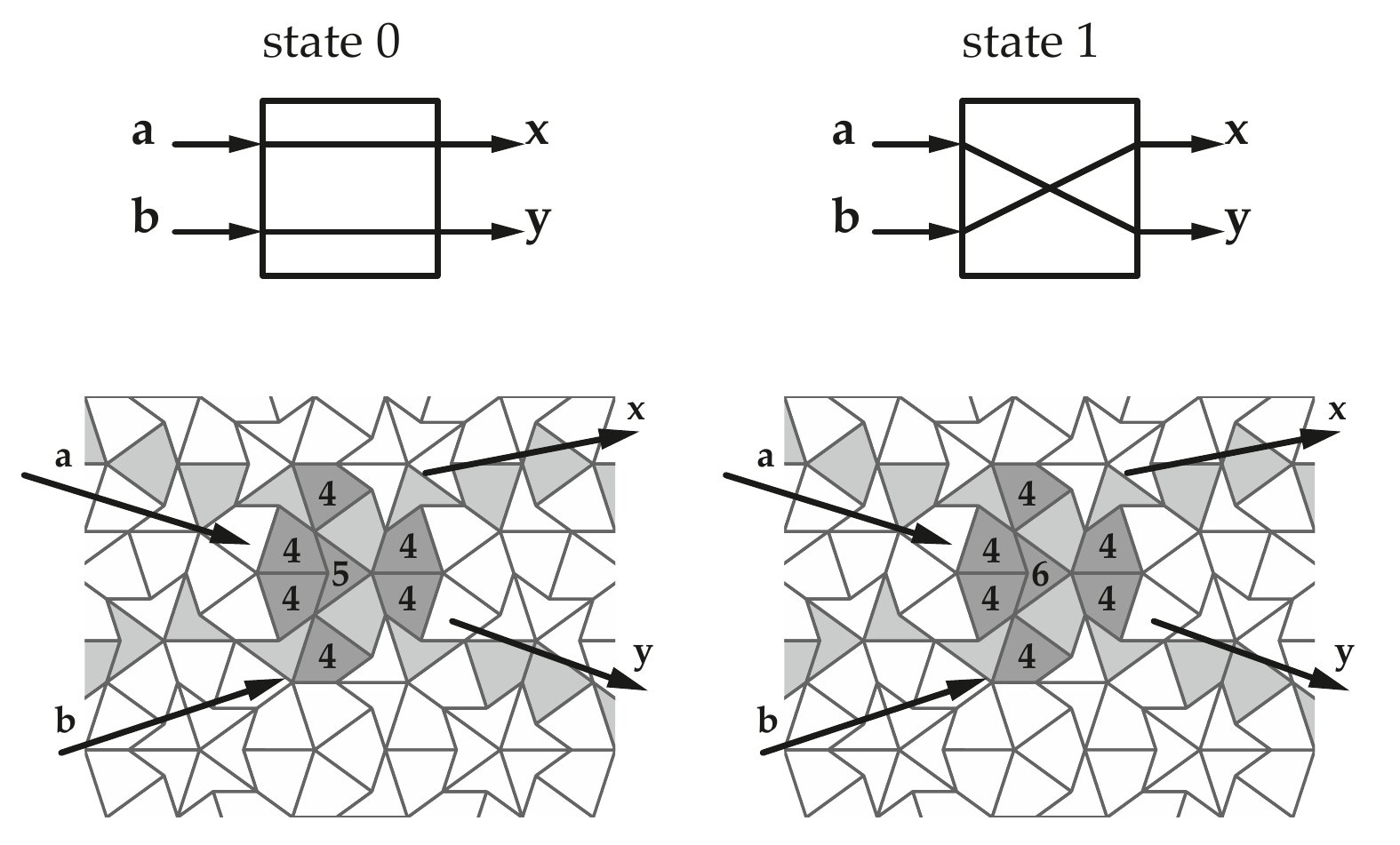}
    \caption{Logical element 2-17 in states 0 (left) and 1 (right). White cells in state 0, light gray cells in state 1.}
    \label{fig:2-2-17}
  \end{center}
\end{figure}

\begin{figure}[htbp]
  \begin{center}
    \includegraphics[width=0.8\linewidth]{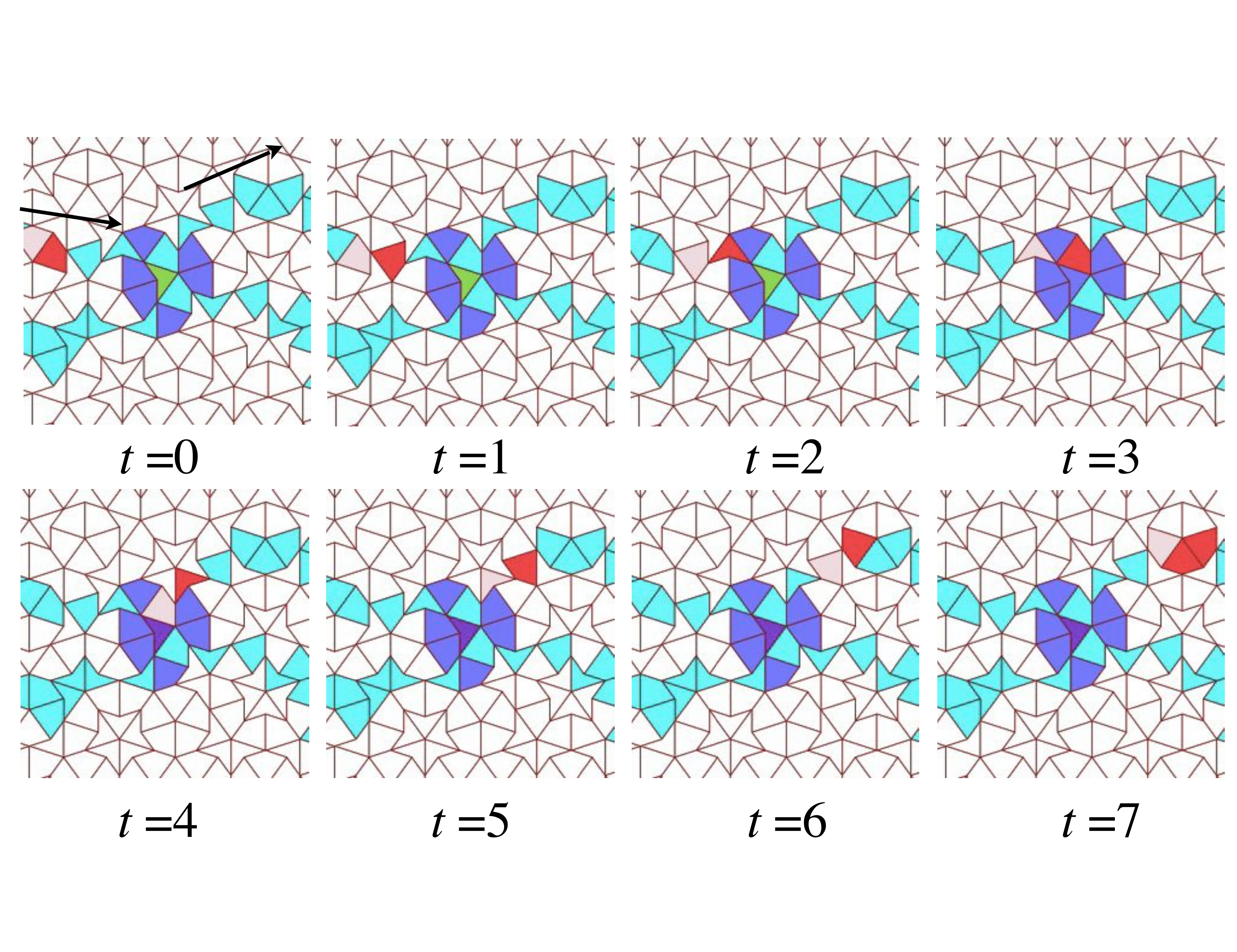}
    \caption{Element 2-17: a transition from state 0, input $a$ to state 1, output $x$.}
    \label{fig:2-17-0b-1x}
  \end{center}
\end{figure}

Similarly, Figures~\ref{fig:2-2-17} and \ref{fig:2-17-0b-1x} illustrate the pattern used to simulate element 2-17.

Because we can control the environment in which the widget evolves, we can precisely set the local transitions up to behave as intended.

\begin{figure}[htbp]
  \begin{center}
    \includegraphics[width=0.5\linewidth]{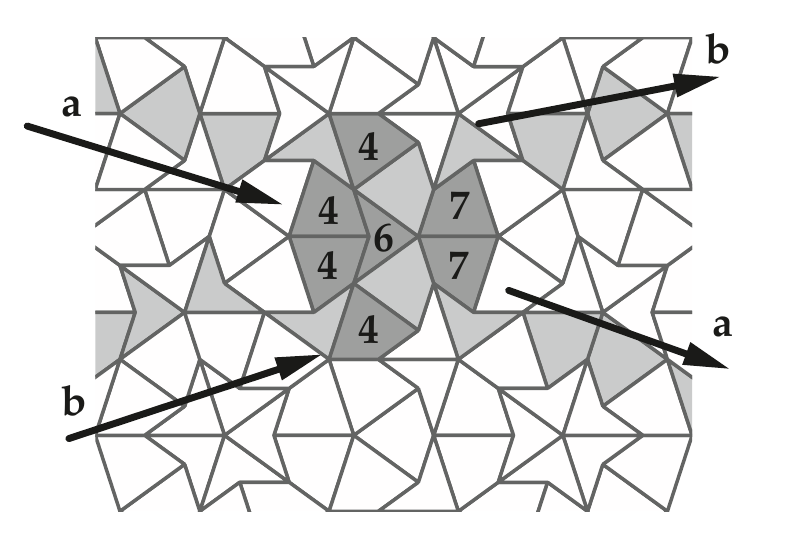}
    \caption{A crossing element. White cells in state 0, light gray cells in state 1.}
    \label{fig:crossing}
  \end{center}
\end{figure}

\begin{figure}[htbp]
  \begin{center}
    \includegraphics[width=0.8\linewidth]{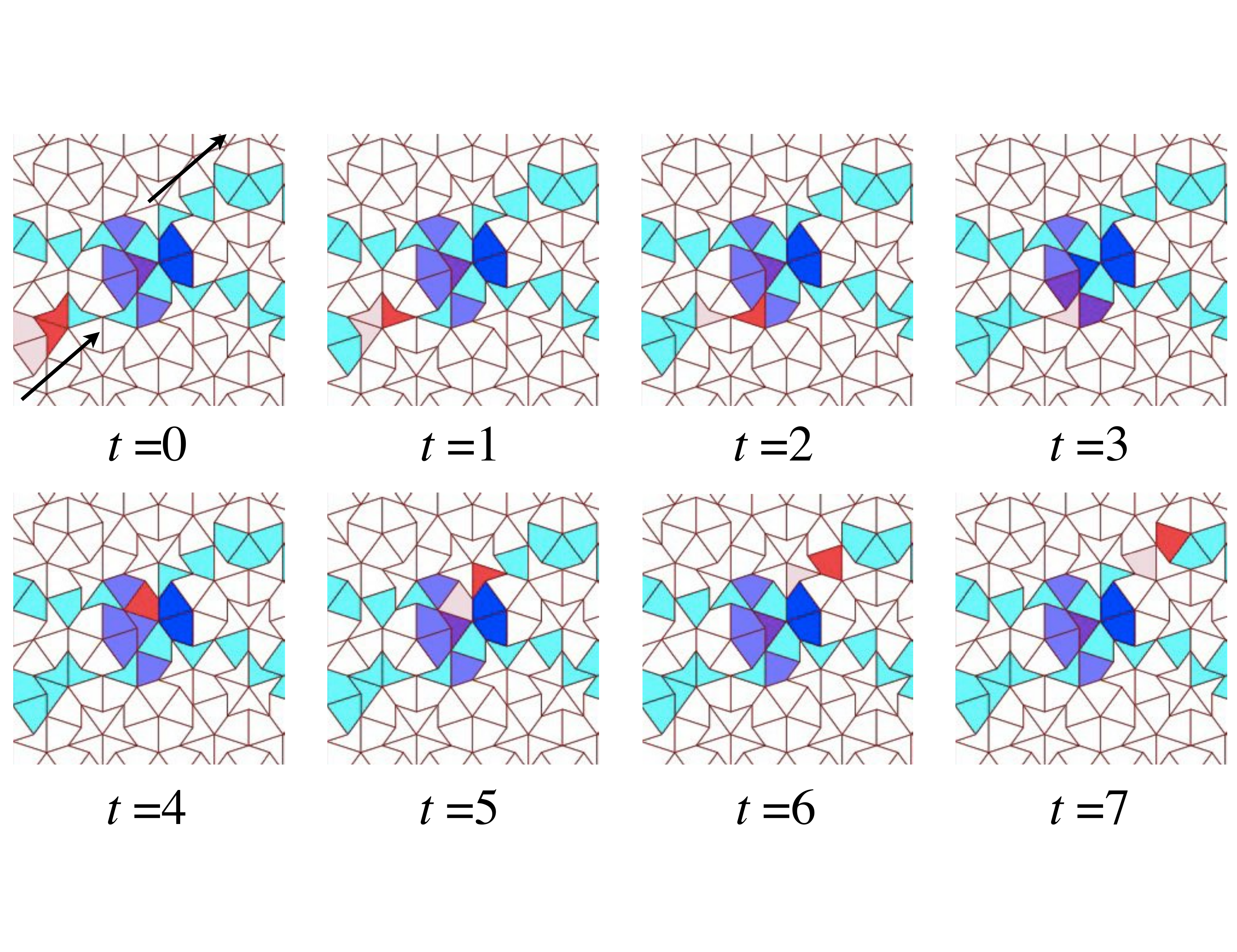}
    \caption{A transtion of a crossing element.}
    \label{fig:crossing-evolution}
  \end{center}
\end{figure}

Finally, in order to be able to connect all the elements freely, it is necessary to add the possibility of crossing wires. This is achieved by a special widget illustrated by Figure~\ref{fig:crossing} and~\ref{fig:crossing-evolution} . This last element is very similar to element 2-17 in its internal state 1, the only difference being that it should not change state when traversed by a signal.

Table~\ref{tbl:rule} gives a full description of the local transition rule of the automaton.

\begin{center}
\begin{table}[htbp]
\[
\begin{array}{| l @{\hspace{.2cm}} l @{\quad} | 
	@{\quad} l @{\hspace{.2cm}} l @{\quad} |
	@{\quad} l @{\hspace{.2cm}} l |}
\hline
0, & *,1,0,1,0,2,0,0 \rightarrow 7 &
1, & *,*,0,1,0,0,1,0 \rightarrow 3 &
1, & *,2,1,1,2,0,0,0 \rightarrow 1 \\
1, & *,1,1,1,0,0,0,0 \rightarrow 1 &
1, & *,2,0,1,4,1,0,0 \rightarrow 3 &
1, & *,1,0,1,3,0,0,0 \rightarrow 1 \\
1, & *,1,0,1,4,0,0,0 \rightarrow 1 &
1, & *,2,0,1,3,0,0,0 \rightarrow 1 &
1, & *,2,0,1,4,0,0,0 \rightarrow 1 \\
1, & *,2,0,2,1,1,0,1 \rightarrow 1 &
1, & *,1,2,1,1,1,0,1 \rightarrow 1 &
1, & *,*,*,1,*,*,*,* \rightarrow 3 \\
1, & *,*,*,2,*,*,*,* \rightarrow 3 &
1, & *,*,*,3,*,*,*,* \rightarrow 3 &
1, & *,*,*,*,2,0,2,1 \rightarrow 1 \\
1, & *,*,*,*,4,0,*,1 \rightarrow 3 &
1, & *,2,0,0,1,1,1,0 \rightarrow 6 &
1, & *,2,0,0,1,0,2,0 \rightarrow 5 \\
1, & *,1,0,0,1,0,1,0 \rightarrow 1 &
1, & *,2,0,0,1,0,1,0 \rightarrow 1 &
1, & *,2,0,0,0,0,1,0 \rightarrow 3 \\
1, & *,3,0,0,1,1,1,2 \rightarrow 6 &
1, & *,4,0,0,0,1,1,1 \rightarrow 6 &
1, & *,2,0,0,1,0,2,1 \rightarrow 5 \\
1, & *,2,0,0,0,0,1,1 \rightarrow 3 &
1, & *,2,0,0,1,2,1,2 \rightarrow 6 &
1, & *,1,0,0,0,0,3,1 \rightarrow 5 \\
2, & *,3,0,1,2,0,0,0 \rightarrow 6 &
2, & *,*,*,*,*,*,*,* \rightarrow 1 &
3, & *,*,*,*,*,*,*,* \rightarrow 2 \\
4, & *,*,0,1,*,0,1,0 \rightarrow 6 &
4, & *,3,0,1,6,0,0,0 \rightarrow 5 &
4, & *,2,1,1,6,0,0,0 \rightarrow 6 \\
4, & *,2,0,2,1,1,0,1 \rightarrow 6 &
5, & *,3,0,1,6,0,0,0 \rightarrow 4 &
5, & *,1,0,0,0,0,3,1 \rightarrow 1 \\
5, & *,4,0,0,0,1,1,1 \rightarrow 6 &
5, & *,2,0,0,1,0,2,1 \rightarrow 1 &
5, & *,3,1,0,1,1,0,2 \rightarrow 1 \\
6, & *,3,0,1,6,0,0,0 \rightarrow 7 &
6, & *,*,1,0,0,0,1,1 \rightarrow 4 &
6, & *,*,1,0,1,0,1,1 \rightarrow 4 \\
6, & *,2,0,0,1,1,2,2 \rightarrow 1 &
6, & *,3,0,0,0,1,2,1 \rightarrow 5 &
6, & *,2,0,0,1,1,2,1 \rightarrow 4 \\
6, & *,3,0,0,1,0,2,1 \rightarrow 4 &
7, & *,3,1,0,4,0,2,0 \rightarrow 4 &
7, & *,0,1,1,0,2,0,0 \rightarrow 0. \\
\hline
\end{array}
\]
\label{tbl:rule}
\caption{Local transition rule of the 8-states universal cellular automaton.\newline The symbol `*' is a wildcard that matches any value. In case of multiple matches, the first matching rule is applied (left to right, top to bottom).}
\end{table}
\end{center}

\subsection{Simulation of a Turing Machine}

It has been shown in~\cite{Morita10} that it is possible to create a circuit that simulates the behavior of a Turing machine cell by arranging eleven copies of a simple logical element with one-bit memory (called \emph{rotary element}). In order to simulate a Turing machine it is then sufficient to connect an infinite line of such circuits. The resulting simulation only requires one active signal to move through the whole infinite circuit.

\begin{figure}[htbp]
	\begin{center}
		\includegraphics[width=\linewidth]{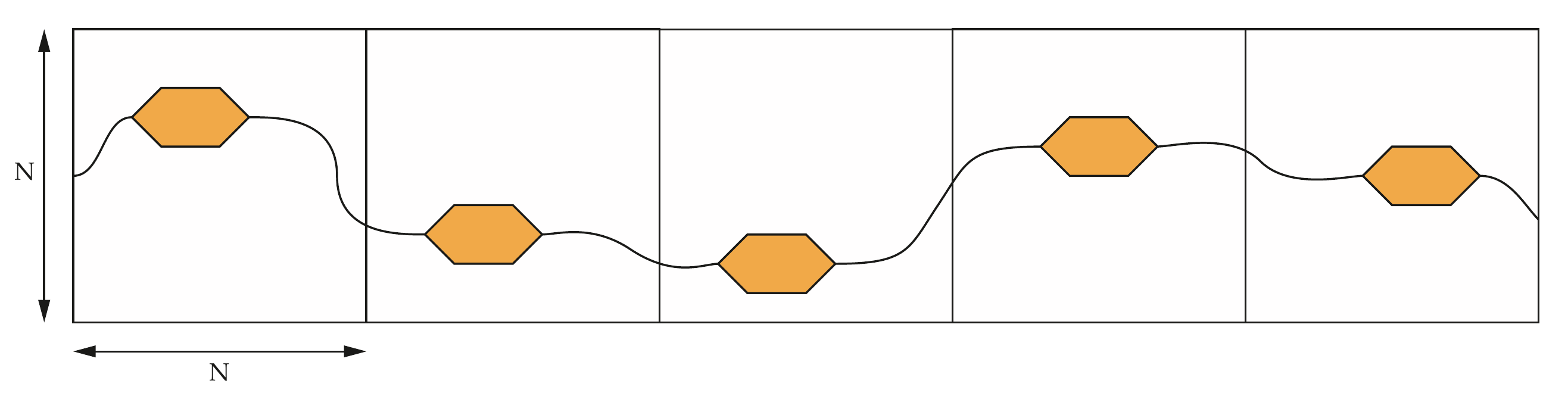}
	\caption{How to simulate the behavior of a Turing machine with copies of a circuit that simulates the behavior of one tape cell.}
	\label{fig:Turing}
	\end{center}
\end{figure}

The rotary element itself can be simulated by elements 2-3 and 2-17 but arranging an infinite line of copies of a given pattern cannot be done periodically on Penrose's tiling. However, since every tiling by Penrose tiles is quasi-periodic, if we can embed a circuit that simulates a Turing machine cell on a certain pattern of the tiling, there exists a constant $N$ such that on every square portion of the plane of size $N\times N$ the pattern appears. This means that if we consider an infinite stripe of the plane divided in squares of size $N\times N$, we can put a Turing machine cell in each square and connect them into an infinite line as illustrated by Figure~\ref{fig:Turing}.

\section{Conclusion}

Despite working on an aperiodic structure with heterogeneous neighborhoods, the cellular automaton that we have described can simulate any logical circuit and even simulate the evolution of a Turing machine, if the infinite circuit is correctly embedded on the tiling (it can be, but in an aperiodic manner). It can work on any valid tiling of the plane by the \emph{kite} and \emph{dart} tiles because all such tilings contain the same finite patterns so any finite configuration on one can be inserted in another.

The number of states of the automaton is likely not minimal for a universal CA. It seems likely that there exist ways to simulate the simple logical elements (or even a different universal combination) with fewer states but possibly larger widgets. It is however very difficult to optimize these constructions as many different environments have to be considered in the search.

Intrinsic universality (the ability to simulate any other cellular automaton) should also be investigated. In that case however, a single-signal asynchronous system cannot work but it might be possible by arranging infinitely many circuits in similar environments, each simulating the behavior of a cell, and connecting them with wires of fixed length to avoid synchronization problems.

\subsubsection*{Acknowledgments.}
Katsunobu Imai gratefully acknowledges the support of the Japan Society for the 
Promotion of Science and the Grant-in-Aid for Scientific Research (C) 22500015.


\begin{thebibliography}{9}
\providecommand{\urlalt}[2]{\href{#1}{#2}}
\providecommand{\doi}[1]{doi:\urlalt{http://dx.doi.org/#1}{#1}}


\bibitem{CR05} Chidyagwai, P., Reiter, CA.: 
  A local cellular model for growth on quasicrystals. 
  \textit{Chaos, Solutions and Fractals} \textbf{24} (2005) 803--812, \doi{10.1016/j.chaos.2004.09.092}.

\bibitem{Dewdney89}  Dewdney, A.K.: 
  Computer recreations: a cellular universe of debris, droplets, defects and demons. 
  \textit{Scientific American} \textbf{261}, August (1989) 102-105.

\bibitem{Gardner70} Gardner, M.:
Mathematical Games - The fantastic combinations of John Conway's new solitaire game ``life'' (1970) 223. pp. 120–123

\bibitem{Gardner77} Gardner, M.:
Extraordinary nonperiodic tiling that enriches the theory of tiles. \textit{Mathematical Games, Scientific American}, January, 1977, p. 110-121.

\bibitem{Griffeath} Griffeath, D.: 
  A CA run on Penrose Tiles, \\
  http://psoup.math.wisc.edu/archive/recipe39.html. 

\bibitem{Fisch90} Fisch, R.: 
  Cyclic cellular automata and related processes. 
  \textit{Physica D} \textbf{45} (1990) 19--25, \doi{10.1016/0167-2789(90)90170-T}.

\bibitem{IIM06} Imai, K., Iwamoto, C., and Morita, K.: 
  A Five-State von Neumann Neighbor Universal Hyperbolic Cellular Automaton. 
  \textit{Journal of Cellular Automata} \textbf{1} 4 (2006) 275--297.

\bibitem{LPAM08} Lee, J., Peper, F., Adachi, S., and Morita, K.:  
  An asynchronous cellular automaton implementing 2-state 2-input 2-output reversed-twin reversible elements. 
  In \textit{Proc. ACRI 2008, LNCS 5191} (2008) 67--76, \doi{10.1007/978-3-540-79992-4\_9}

\bibitem{MOA11} Morita, K., Ogiro, T., and Alhazov, A.: 
  Non-degenerate 2-state reversible logic elements with three or more symbols are all universal. 
  \textit{Multiple-Valued Logic and Soft Computing} \textbf{18} 1 (2012) 37-54.

\bibitem{Morita10} Morita, K.: 
  Constructing a reversible turing machine by a rotary element, a reversible logic element with memory. 
  \textit{In Hiroshima University Institutional Repository} \\ http://ir.lib.hiroshima-u.ac.jp/00029224.

\bibitem{MM11} Mukai, Y., Morita, K.: 
  Universality of 2-symbol reversible logic elements with memory. 
  \textit{Resume for LA Symposium Summer} (2011), \textbf{S16} 1--15 (in Japanese).

\bibitem{OS08} Owens, N., Stepney, S.: 
  Investigation of the Game of Life cellular automata rules on Penrose tilings: lifetime, ash and oscillator statistics. 
  \textit{Proc. AUTOMATA 2008} (2008) 1--34. 

\bibitem{OS10} Owens, N., Stepney, S.: 
  The Game of Life rules on Penrose tilings: still life and oscillators. 
  In: A. Adamatzky (Ed.) \textit{Game of Life Cellular Automata}, Springer-Verlag London (2010) 331--378. 

\bibitem{Reiter10} Reiter, CA.: 
  Medley of spirals from cyclic cellular automata. 
  \textit{Computers \& Graphics} \textbf{34} (2010) 72--76, \doi{10.1016/j.cag.2009.04.008}.

\bibitem{Weeks} Weeks, E.: 
  Cellular automata on quasicrystals, \\ http://www.physics.emory.edu/\~{}weeks/pics/qvote.html.

\bibitem{Wojtowicz} W\'ojtowicz, M.: 
  Cellular Automata rules lexicon, \\ http://www.mirekw.com/ca/ca\_rules.html.

\end{thebibliography}
\end{document}